\documentclass[%
 aip,
 amsmath,amssymb,
 reprint,%
]{revtex4-1}

\usepackage{graphicx}
\usepackage{dcolumn}
\usepackage{bm}
\usepackage{scalerel}
\usepackage[utf8]{inputenc}
\usepackage[T1]{fontenc}
\usepackage{mathptmx}
\usepackage{xcolor}

\usepackage{epstopdf}

\definecolor{cream}{RGB}{222,217,201}
\pdfoutput=1
\begin{document}
\preprint{AIP/123-QED}

\title{Impurity effects in thermal regelation}

\author{Navaneeth K. Marath}
\affiliation{%
Nordita, Royal Institute of Technology \& Stockholm University, Stockholm 106 91, Sweden
}%

\author{John S. Wettlaufer}
\affiliation{%
Yale University, New Haven, Connecticut 06520,United States
}%
\affiliation{%
Nordita, Royal Institute of Technology \& Stockholm University, Stockholm 106 91, Sweden
}%
\date{\today}

\begin{abstract}{When a particle is placed in a material with a lower bulk melting temperature, intermolecular forces can lead to the existence of a ``premelted'' liquid film of the lower melting temperature material.  Despite the system being below the melting temperatures of both solids, the liquid film is a consequence of thermodynamic equilibrium, controlled by intermolecular, ionic and other interactions.     
An imposed temperature gradient drives the translation of the particle by a process of melting and refreezing known as ``thermal regelation''.  We calculate the rate of regelation of spherical particles surrounded by premelted films that contain ionic impurities. The impurities enhance the rate of motion thereby influencing the dynamics of single particles and distributions of particles, which we describe in addition to the consequences in natural and technological settings.  } \\

\end{abstract}

\maketitle


\section{Introduction}\label{sec:intro}

Premelted liquid films can separate the surface of a solid from a foreign substrate at temperatures below the solid's bulk melting temperature.  The solid melts against the substrate in order to minimize the free energy of the solid-liquid-substrate system\cite{dash2006physics}. Thorough reviews \cite{dash1995premelting,dash2006physics,JSWARFM} discuss the theory of premelting and its consequences in a swath of natural and technological settings. One of the consequences of premelting is the translation of a foreign particle through a solid when subjected to a temperature gradient.  In this process of {\em thermal regelation} the solid melts at the warmer side of the particle and the fluid flows to the colder side and refreezes thereby facilitating the translation. Evidently Gilpin\cite{GILPIN1979235} first modeled thermal regelation, which was later discussed in the context of interfacial melting\cite{worster1999fluid} and frost heave\cite{rempel2004premelting}. 
In most settings impurities are present in the films surrounding particles, but the rate of regelation has only been calculated for particles surrounded by pure films. 
The effect of impurities on the premelting of ice has been calculated for a wide range of planar substrates \cite{wettlaufer1999}.  By combining the key ingredients of pure and impure systems \cite{worster1999fluid,wettlaufer1999,rempel2004premelting,hansen2010theory}, we analyze the effects of impurities on the rate of thermal regelation of spherical particles and discuss their implications  for environmental and engineering problems, by treating specific materials, in particular ice.  

In \S \ref{sec:theory}, we give the theory for thermal regelation of spherical particles embedded in a solid. First, we describe how impurities in the premelted films that surround the particles control the film thickness. Then, we calculate the translational velocity and displacement of the particle, when it is subjected to a temperature gradient. Lastly, we develop the theory to understand the combined effects of thermal regelation and diffusion on the particle motion.  In \S \ref{sec:Results}, we describe the displacement of particles of different sizes and impurity concentrations for various temperature gradients.  The examples we provide are motivated by the dating of ice cores used in paleoclimate research and the manipulation of composite materials, as described in \S \ref{sec:app}.  
 We conclude in \S \ref{sec:conclusions}.

\section{Theory}\label{sec:theory}
\begin{figure}
\centering
 \includegraphics[height=8cm]{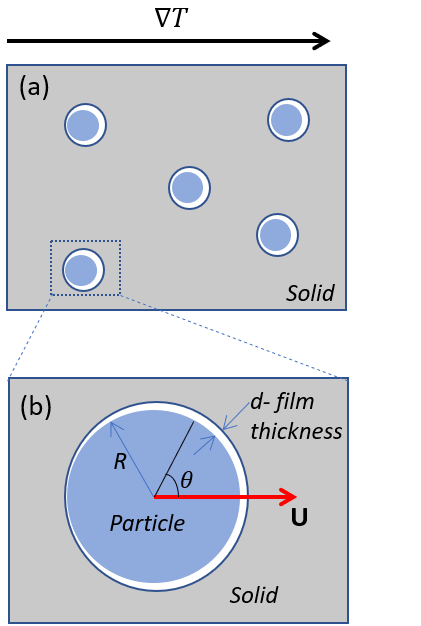}
 \caption{(a) shows particles embedded in a solid subjected to a temperature gradient of $\nabla T$ and 1(b) shows a zoomed view of the particle that is enclosed by the box in 1(a). The radius and the translational velocity of the particle are denoted by $R$ and $\mathbf{U}$, respectively. The premelted film that separates the solid from the particle is of thickness $d$.}
 \label{figure1}
 \vspace{-0.25 cm}
\end{figure}

We analyse the translation of a spherical particle of radius $R$ that is surrounded by a solid, which premelts against the particle  forming a thin film of liquid of thickness $d$, as shown in figure \ref{figure1}(b). 
At temperatures below the solid's bulk melting temperature, $T_m$, the film's thickness depends on the nature of the intermolecular and Coulombic forces operative in the system and is a function of the impurity concentration and the degree of undercooling \cite{wettlaufer1999,hansen2010theory}. For sufficiently large concentrations \cite{wettlaufer1999,hansen2010theory,dash2006physics}, the thickness of the premelted film on a flat substrate ($R\rightarrow \infty$) is dominated by colligative effects and is given by  
\begin{align}
 d= \frac{R_g T_m^2 N_i}{\rho_l q_m \Delta T}\label{equation7}.
\end{align}
Here $R_g$ is the universal gas constant, $\rho_l$ is the molar density of the liquid, $\Delta T = T_m-T$ is the undercooling, with $T$ is the temperature of the solid/liquid/substrate system, $q_m$ is the latent heat of melting per mole of the solid and $N_i$ is the number of moles of impurity per unit surface area of the substrate. 
The concentrations and the temperatures in which this linear colligative-type relationship between $d$, $N_i$ and $\Delta T$ holds, depends on the nature of the materials in the system \cite{wettlaufer1999}. 

Firstly, for spheres of radius $R$, so long as $d\ll R$, the curvature of the solid-liquid interface is approximately the same as that of the particle-liquid interface.
Secondly, for sufficiently large impurity concentrations and particles, then the Gibbs-Thomson effect does not control the film thickness \cite{hansen2010theory}.   This requires that $\Delta T > 2 T_m\gamma_{sl}/R \rho_s q_m$, so that for micron sized particles in ice, where $\gamma_{sl}=$0.033 $J/m^2$ is the solid-liquid interfacial free energy,  $T_m=$273.15 $K$, and $\rho_s q_m\approx$ 334$\times 10^6 J/m^3$, and thus $\Delta T > 0.05 K$.  Therefore, unless the system is within 50 $mK$ of the bulk melting point, Eq.\ref{equation7} determines the film thickness for high concentrations and particles larger than about a micron, which is the situation we treat here. 

\subsection{Rate of thermal regelation}\label{sec:thermal}
 The pressure in the solid ($p_s$) will be different from that in the liquid ($p_l$) and the difference is calculated using the Gibbs-Duhem equation
\cite{callen1998thermodynamics} 
 and is given by $p_s-p_l\approx \rho_s q_m \Delta T/T_m$. This pressure difference creates a force acting on the particle given by
\begin{align}
\mathbf{F}_p\approx-\displaystyle \int_{\mathbf{S}}  \displaystyle\frac{\rho_s q_m \Delta T}{T_m} \mathbf{n} dA=\int_{V}  \frac{\rho_s q_m \nabla T}{T_m} dV=\frac{m_s q_m \langle\nabla T\rangle}{T_m} \label{equation8},
\end{align}
where $\mathbf{n}$ is the unit vector normal to the surface ($\mathbf{S}$) of the particle-film system,  $V$ is the volume, $m_s$ is the mass of the displaced host solid, and $\langle\nabla T\rangle$ is the temperature gradient averaged over $V$. We assume that the materials have the same thermal conductivities and the temperature gradient is a constant vector, which implies $\langle\nabla T\rangle=\nabla T$. The force in Eq.\ref{equation8} is a ``thermodynamic buoyancy'' force \cite{rempel2001interfacial} and it pushes the particle in the direction of the gradient--towards the warmer side of the particle. As the particle moves, the thickness of the film must obey Eq.\ref{equation7}$\,$, with larger thicknesses at larger temperatures.  Therefore,  the solid melts on the warm side, and the melt flows through the film to the colder side where it refreezes. The flow is driven by a thermomolecular pressure gradient \cite{rempel2001interfacial}, or temperature gradient induced liquid pressure gradient, and it will exert a hydrodynamic force on the particle governed by lubrication theory \cite{batchelor2000introduction}.  At a polar angle $\theta$ (figure \ref{figure1}(b)) the volume flux from lubrication theory is equated to the particle motion as
\begin{align}
 -\pi (R  sin \theta)^2 U = \frac{\pi d^3 \sin \theta}{6 \mu} \frac{d p_l}{d\theta}\label{equation9},
\end{align}
where $U$ is the magnitude of the translational velocity of the particle and $\mu$ is the viscosity of the liquid. 
Substituting the thickness from Eq.\ref{equation7} into Eq.\ref{equation9} and integrating the latter with respect to $\theta$ we obtain the liquid pressure \cite{rempel2004premelting}. Hence, the hydrodynamic force is given by $\mathbf{F}_l=-\int_{\mathbf{S}} p_l \mathbf{n} dA $, and integration yields
\begin{align}
 \mathbf{F}_l=-\frac{8\pi \mu R^4 }{d_0^3}\mathbf{U},
\end{align}
where $d_0$ is the thickness of the film at the equator of the particle, when its axis is parallel to the temperature gradient. The expression for the hydrodynamic force is the same as that derived for the case of a premelted film without impurities \cite{worster1999fluid,rempel2004premelting}; however, here the relationship between $d_0$ and the undercooling $\Delta T$ is given by Eq.\ref{equation7}. Equating the hydrodynamic and thermodynamic buoyancy forces we obtain the translational velocity of the particle as
\begin{align}
 \mathbf{U}=\frac{\rho_s q_m \nabla T d_0^3}{6 \mu R T_m}\label{equation11}, 
\end{align}
showing that the particle translates parallel to the temperature gradient.

\subsection{Particle displacement by thermal regelation}\label{sec:displacement}
Substituting the thickness from Eq.\ref{equation7} into Eq.\ref{equation11} we can rewrite the latter as
\begin{align}
 U(z_p)=\frac{d z_p}{dt}= \frac{A_3}{(A_1-A_2 z_p)^3}\label{equation12},
\end{align}
where ${z}_p$ is the position of the particle measured parallel to the temperature gradient, with $z_p=0$ at $t=0$, $A_1=\rho_l q_m (T_m-T_{int})/T_m$, $A_2=\rho_l q_m \lvert\nabla T\rvert/T_m$, $A_3=\rho_s q_m\lvert \nabla T\rvert (R_g T_m N_i)^3/6\mu R T_m$, and $T_{int}$ is the temperature at the initial position of the particle. Integration of Eq.\ref{equation12} leads to a quartic equation in $z_p$ whose solution is
\begin{align}
 z_p= \frac{A_1-(A_1^4-4 A_2 A_3 t)^{1/4}}{A_2}\label{equation13},
\end{align}
which gives the net displacement of the particle at time $t$.  When $z_p\ll A_1/A_2$, then the displacement grows linearly with time, viz., $z_p\sim \frac{A_3}{A_1^3} t$,  followed by power law growth.

\subsection{Particle diffusion}\label{sec:diffusion}
Owing to the thermal fluctuations in the premelted film a spherical particle can execute diffusive motion through the host solid. The premelting-controlled diffusivity of the particle is given by $\mathbf{D}=(3d_0^3/4R^3)\mathbf{D_s}$, where $\mathbf{D_s}=k_b T \mathbf{I} /6\pi \mu R $ is the Stokes-Einstein diffusivity of the particle, $\mathbf{I}$ is the second-order identity tensor and $k_b$ is Boltzmann's constant\cite{peppin2009onsager}. In a Cartesian coordinate system, whose $z$-axis is parallel to the temperature gradient with $T(z=0)=T_m$, the premelting-controlled diffusivity of the particle located at $\mathbf{x}=(x,y,z)$ is given by 
\begin{align}
 \mathbf{D}=\frac{(R_g T_m N_i)^3 }{8 \pi \mu R^4{A_2}^3  }\frac{ k_b T_{m}}{z^3}\mathbf{I}= D(z) \mathbf{I} \label{equationD},
\end{align}
where we have used thickness from Eq.\ref{equation7}, with $\Delta T= z \lvert\nabla T\rvert$, and $A_2$ is the thermomolecular pressure coefficient times $\lvert \nabla T\rvert $, as defined following Eq.\ref{equation12}. 

\subsection{Combining thermal regelation and diffusion}\label{sec:fpe}

The particle translates due to thermal regelation, and using Eq.\ref{equation11}, its velocity can be written in the Cartesian coordinate system as
\begin{align}
 \mathbf{U}=-\frac{A_3}{A_2^3}  \frac{1}{z^3}\mathbf{\hat{i}}=U(z)\mathbf{\hat{i}}\label{equation15},
\end{align}
where $A_3$ is defined below Eq.\ref{equation12}.  Hence, we combine the effects of regelation (Eq. \ref{equation15}) and diffusion (Eq. \ref{equationD}) in a statistical mechanical treatment of particles in the dilute limit as follows.  The probability of finding a particle at $\mathbf{x}=(x,y,z)$ at time $t$ is given by the probability density $f(\mathbf{x}, t)$, which is governed by the following Fokker-Planck-like equation
\begin{align}
 \frac{\partial f}{\partial t}+  \frac{\partial }{\partial z}[U(z) f]= \frac{\partial}{\partial z}\left[D(z)\frac{\partial f}{\partial z}\right] +D(z)\left[ \frac{\partial^2 }{\partial x^2}+\frac{\partial^2 }{\partial y^2}\right] f \label{equation16}.
 \end{align}
 Since the translational velocity and the diffusivity have a non-trivial dependence on $z$, a complete understanding of this theory requires a numerical treatment.  However, rewriting Eq. \ref{equation16} as 
 \begin{align}
 \frac{\partial f}{\partial t}+ \left[U(z)- \frac{\partial D(z)}{\partial z}\right]\frac{\partial f}{\partial z} = -\frac{\partial U(z)}{\partial z} f +D(z)\nabla^2 f \label{equation17}, 
 \end{align}
 facilitates an analytical interpretation using a ``P\'eclet function'' defined as 
\begin{align}
Pe(z)=\frac{\left[U(z)-\frac{\partial D(z)}{\partial z}\right]}{D(z)}L_z ,\label{equation18}
\end{align}
where $L_z$ is the characteristic length scale in the $z$-direction. Now, we can solve Eq. \ref{equation17} analytically in the large $Pe(z)$ limit, that is when diffusion in the $z$-direction can be neglected.

\section{Results}\label{sec:Results}

Our model system consists of silicon particles embedded in ice.  We use these materials for two main reasons.  Firstly, ice is a material that exhibits all of the phase behavior of general interest in the physical sciences, but with distinct advantages associated with its ready accessibility, transparency and experimental control\cite{dash2006physics}.  Therefore, it acts as a test bed for a broad class of materials processes, and of particular relevance here are composite materials.  Secondly, ice is an essential astro-geophysical material and thus basic processes, such as particle migration in ice cores, are of great contemporary interest, as described in \S \ref{sec:app}.   

\subsection{Thermal regelation alone}\label{sec:regelationresults}
\begin{figure}[ht]
\centering
 \includegraphics[height=6.5cm]{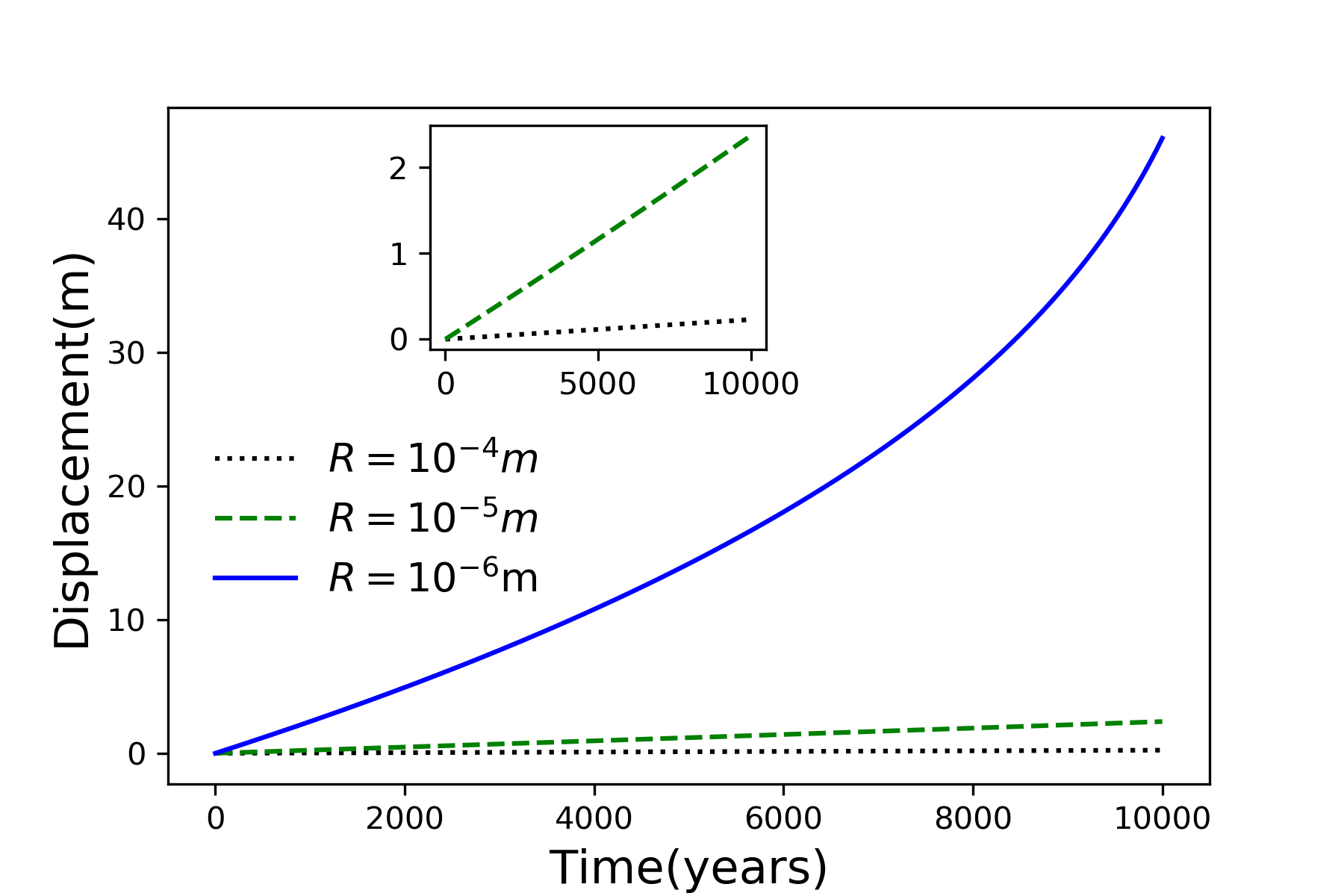}
  \caption{ Displacement versus time for different particles  with $T_m-T_{int}$=10 $K$, $\lvert\nabla T\rvert$= 0.1 $K/m$ and $N_i$=100$ \,\mu M/m^2$. The inset is an expanded view of the plot for the particles with $R=10^{-4}$ and $10^{-5}$ $m$ .}
 \label{figure2}
  \vspace{-0.25 cm}
\end{figure}

\begin{figure}[ht]
\centering
 \includegraphics[height=6.5cm]{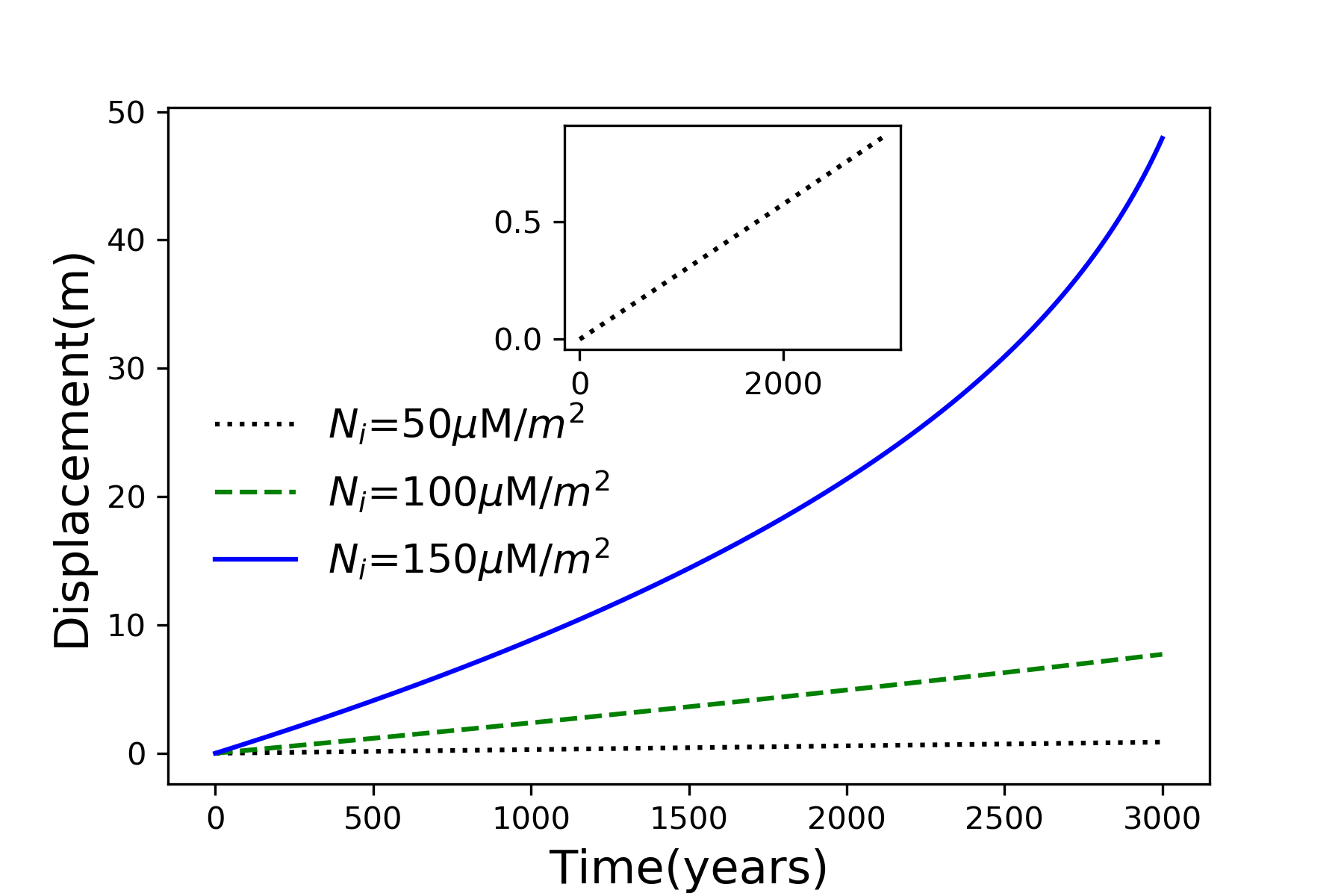} 
 \caption{ Displacement versus time for a particle of radius $10^{-6}m$ at three different concentrations with $T_m-T_{int}$=10 $K$ and $\lvert\nabla T\rvert$= 0.1 $K/m$. The inset is an expanded view of the plot for the particle with $N_i$= 50 $ \,\mu M/m^2$.}
 \label{figure3}
  \vspace{-0.25 cm}
\end{figure}

\begin{figure}[ht]
\centering
 \includegraphics[height=6.5cm]{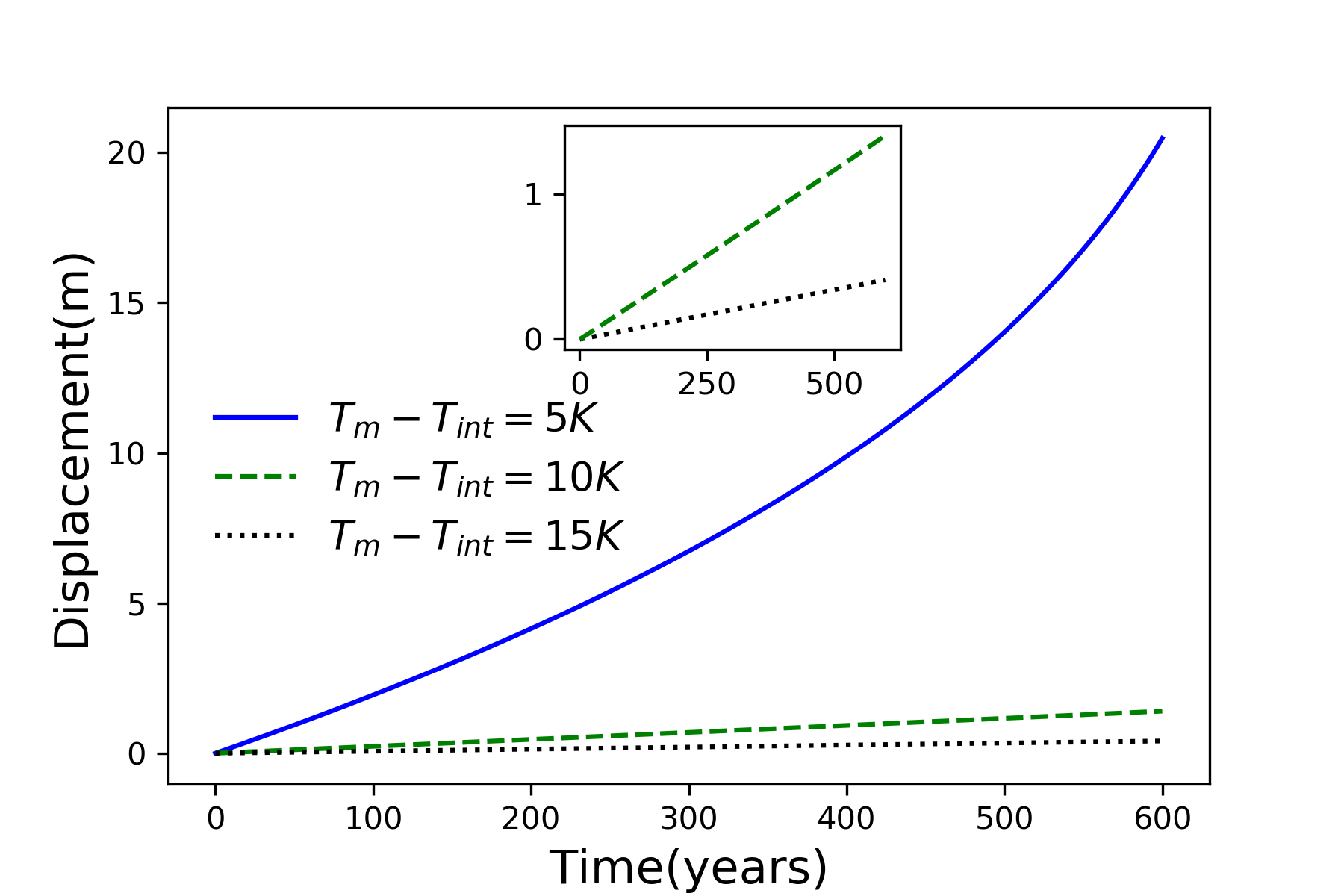}
  \caption{Displacement versus time  for a particle of radius $10^{-6}m$ starting from three different undercoolings with $N_i$= 100 $\,\mu M/m^2$ and $\lvert\nabla T\rvert$= 0.1 $K/m$. The inset is an expanded view for the particles starting at $T_m-T_{int}$= 10 $K$ and 15 $K$.}
 \label{figure4}
  \vspace{-0.25 cm}
\end{figure}

Using Eq.\ref{equation13}, we plot the displacement of particles of three different sizes versus time in figure \ref{figure2}, with $T_{int}=$263.5 $K$, which corresponds to an undercooling of  10 $K$. We have set the temperature gradient to 0.1 $K/m$, and the concentration of impurities to 100$ \,\mu M/m^2$. The biggest particle considered has the smallest displacement, and in $10^4$ years, the particle of radius $10^{-6}m$ experiences both linear and nonlinear growth in its displacement.
To understand the effects of impurities on thermal regelation, in figure \ref{figure3}, we have plotted the displacement of a particle of radius $10^{-6}m$ at three different concentrations, with $T_{int}=$263.5 $K$ and $\lvert\nabla T\rvert$= 0.1 $K/m$. Eq.\ref{equation11} shows that the translational velocity of the particle is proportional to the thickness of the film at its equator, and that thickness increases with concentration as governed by Eq.\ref{equation7}$\,$, so that the displacement increases with concentration $N_i$. 
In figure \ref{figure4} we show the dependence of the displacement on $T_{int}$ for a particle of radius $10^{-6}m$ starting at three different initial temperatures at a concentration of $N_i$= 100 $\,\mu M/m^2$ and $\lvert\nabla T\rvert$= 0.1 $K/m$. The particle starting at the smallest undercooling experiences the largest displacement because the thickness of the film increases with $T_{int}$. 
 
\begin{figure}
\centering
 \includegraphics[height=6.5cm]{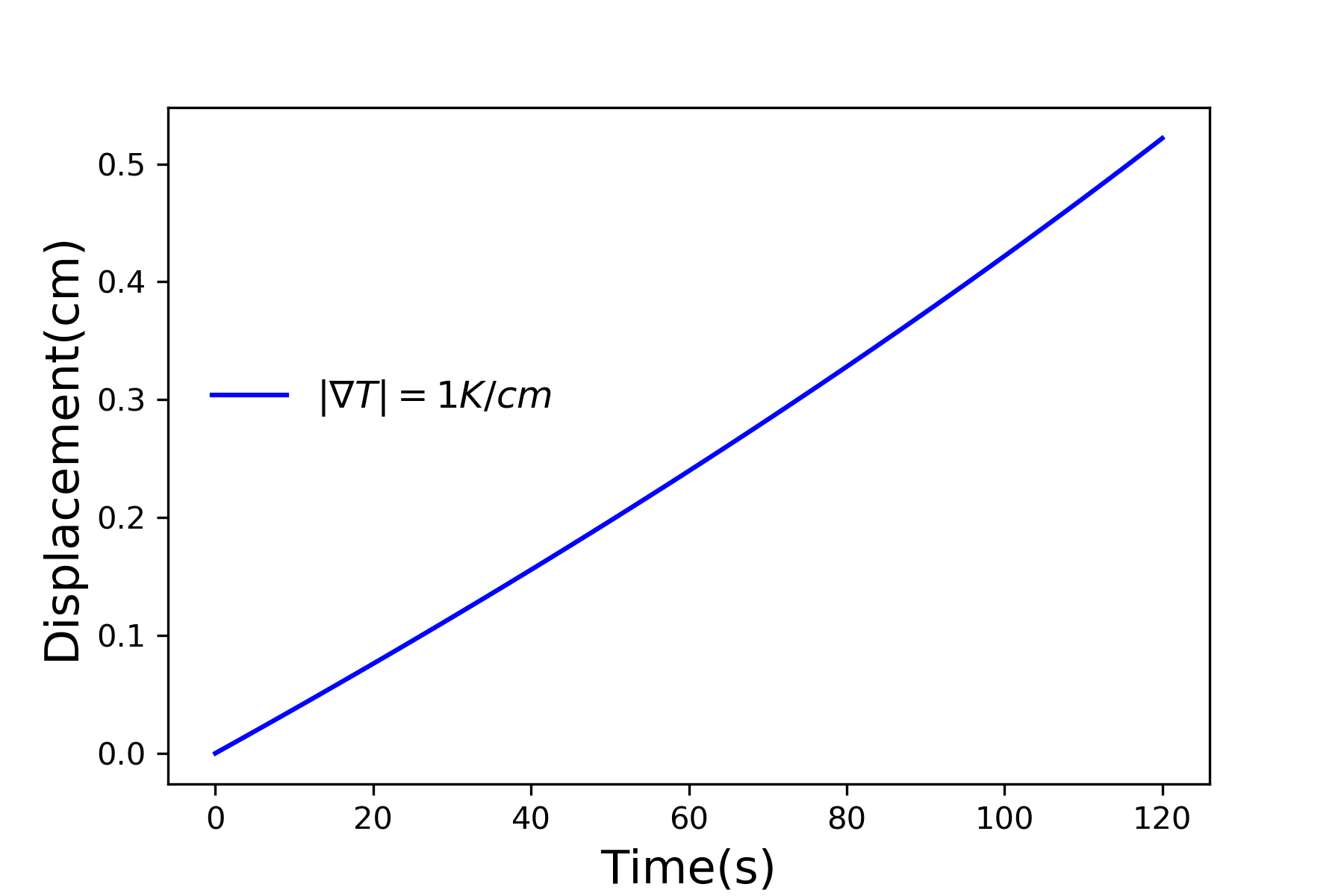}
  \caption{Displacement versus time for a particle of radius $10^{-6} m$ with $N_i$= 400 $ \,\mu M/m^2$, $T_m-T_{int} = 5 K$ and $\lvert\nabla T\rvert$= 1 $K/m$. }
 \label{figure5}
  \vspace{-0.25 cm}
\end{figure}

Figures \ref{figure2}-\ref{figure4} have time scales of the order of years and length scales of the order of meters. In a lab or composite manufacturing setting, the time and length scales are of the order of minutes to hours and centimeters. Such small scales can be achieved by using a larger temperature gradient and in figure \ref{figure5} we have plotted the displacement of a particle of radius $10^{-6}m$, in a temperature gradient of 1 $K/cm$ with $N_i$= 400 $ \,\mu M/m^2$. Clearly, by adjusting the magnitude of the temperature gradient and the concentration of impurities, the motion of a particle inside a solid can be controlled on laboratory scales. 
\subsection{Thermal regelation and diffusion}\label{sec:RDresults}
\begin{figure}
\centering
 \includegraphics[height=5.1cm]{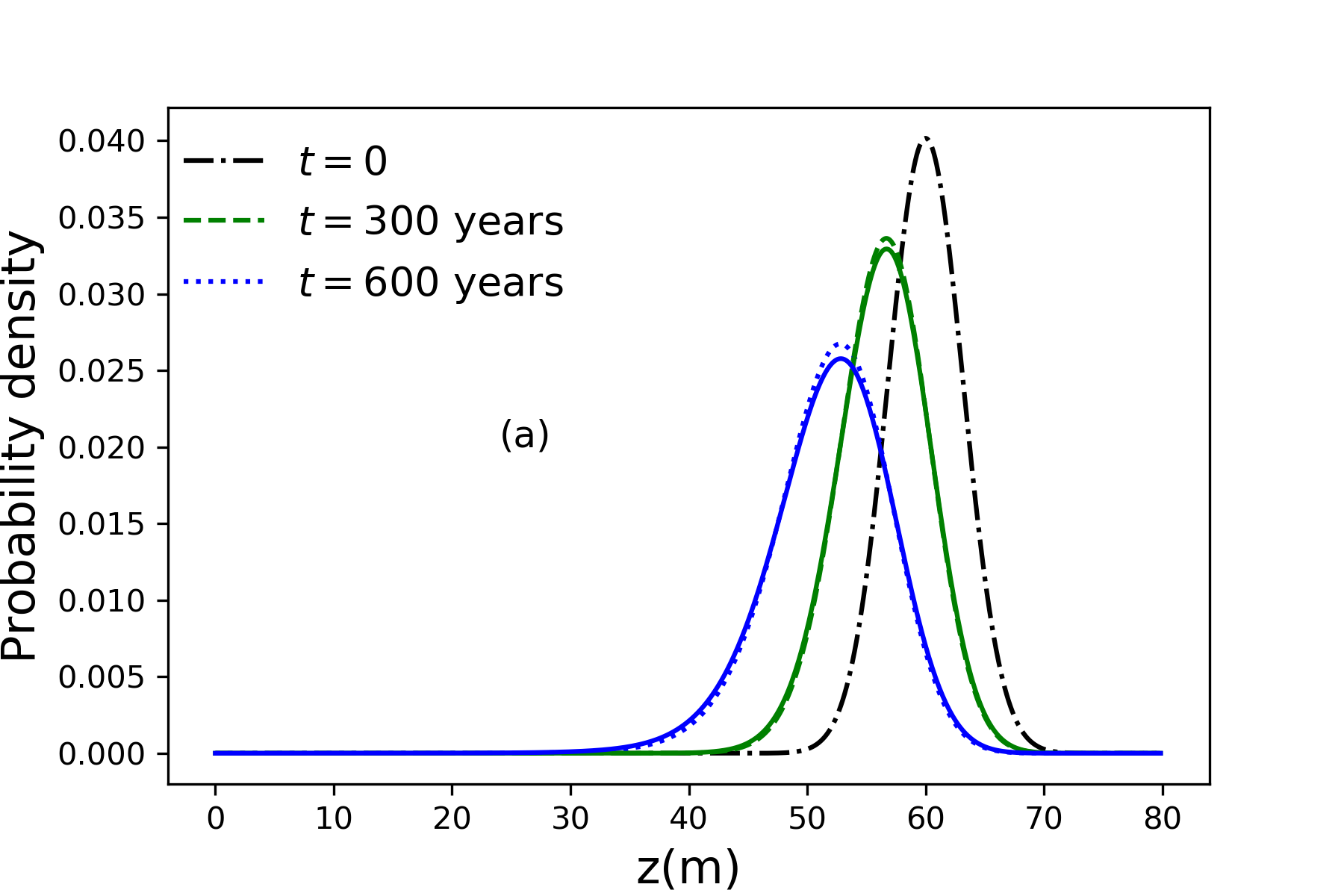}\\
 \includegraphics[height=5.1cm]{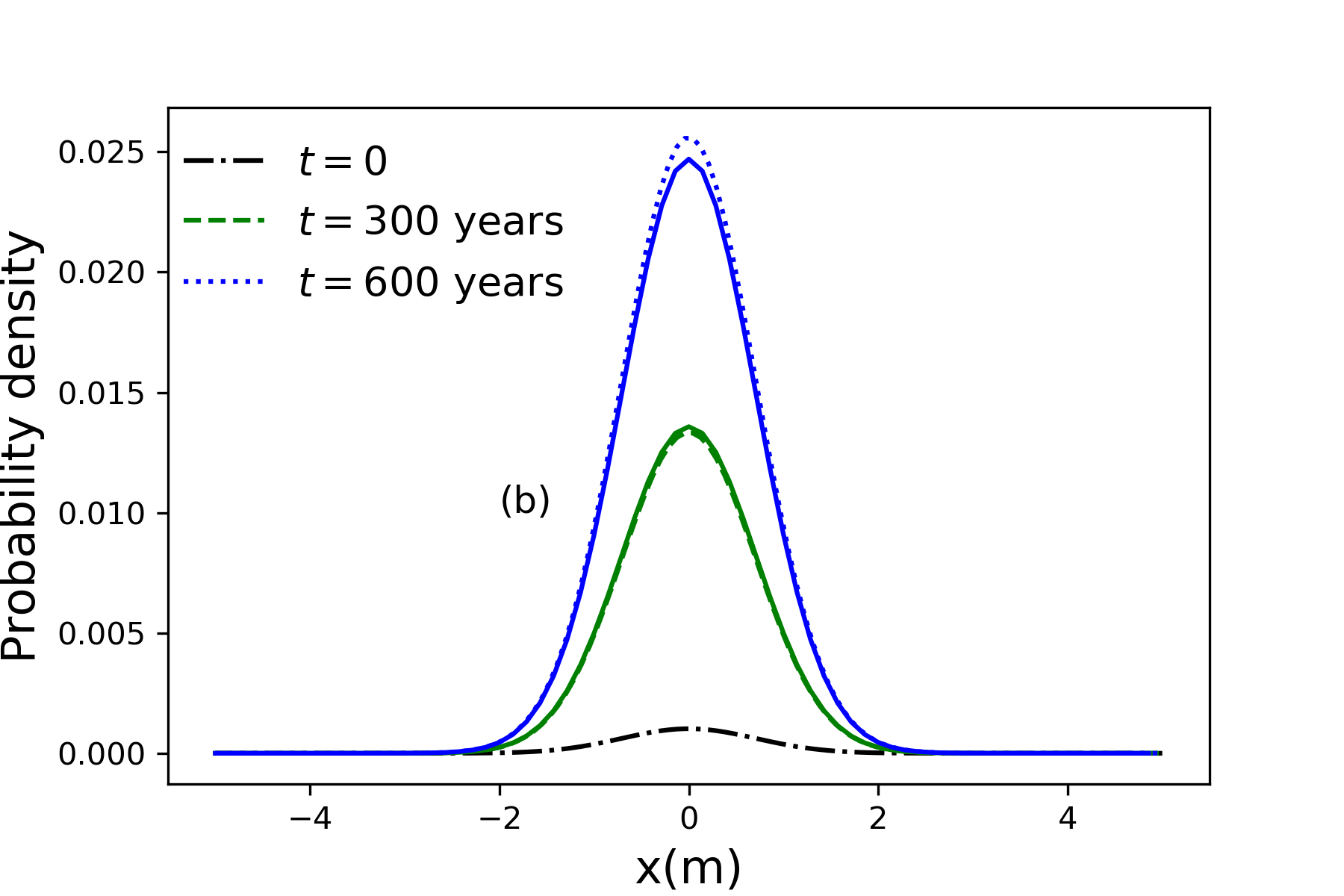}
  \caption{(a) Evolution of the probability density along the $z$ axis at $x=0$ and $y=0$ for particles of radius $10^{-6} m$ with $N_i$= 100 $\,\mu M/m^2$, $\lvert\nabla T\rvert$= 0.1 $K/m$ and $T=T_m$ at $z=0$. (b) Evolution of the probability density along the $x$ axis at $y=0$ and $z=51$.}
 \label{figure6}
  \vspace{-0.25 cm}
\end{figure}

As noted above, thermal fluctuations in the premelted films that surround the particles inside a solid facilitates their diffusion through the solid. As shown in Eq.\ref{equation17}, the combined effects of thermal regelation and diffusion will determine the evolution of the probability density of particles, $f(\mathbf{x}, t)$. We analyze the evolution of $f(\mathbf{x}, t)$ by numerically solving Eq.\ref{equation17}  for micron-sized silicon particles in ice. In figures \ref{figure6} (a) and (b) we have plotted the evolution along the $z$ and $x$ axes, respectively. The initial distribution is assumed to be a three-dimensional Gaussian, $f(\mathbf{x},0)=e^{-x^2-y^2-(z-60)^2/20}/7.926\pi$, peaked at $\mathbf{x}=\{0,0,60\}$. If the characteristic length scale in Eq.\ref{equation18} is set as $1\, m$, then $Pe$ is of the order of $10^8$ for the solution presented in the figures. The numerical solution (solid lines) compares well with the large $Pe$ analytical solution of Eq.\ref{equation17}  (dotted and dashed lines) given by 
\begin{equation}
f(\mathbf{x},t)= \frac{z^3}{z'^{\frac{3}{4}}}\frac{\exp{\left[~-~{\frac{(x^2+y^2)}{1+4 \frac{D(z)}{U(z)}\left(z'^{\frac{1}{4}}-z\right)}}\right]}}{7.926\pi\left[1+4 \frac{D(z)}{U(z)}\left(z'^{\frac{1}{4}}-z\right)\right]} \exp{\left(-\frac{\left[z'^{\frac{1}{4}}-60\right]^2}{20} \right)},
\end{equation}
for this particular initial condition, where $z'=z^4+4A_3\,t/A_2^3 $. In figure \ref{figure6} (a), the advection of the probability density towards higher temperatures ($T=T_m$ at $z=0$) is evident and the decay in the probability density is due to the reaction term ($\propto f$) in Eq.\ref{equation17}. The growth of density in figure \ref{figure6} (b) is a consequence of the advection. Thus, for micron sized silicon particles in ice subjected to a gradient of $0.1 K/m$ with $N_i= 100 \,\mu M/m^2$, the advection and the reaction terms in Eq.\ref{equation17} determine the evolution of the probability density. Although we have illustrated the solution of Eq.\ref{equation17} only for a large P\`eclet function, clearly the solution behavior can alter between the advection-dominated one seen in figure \ref{figure6} and a diffusion-dominated one depending on the temperature gradient, particle size and/or impurity concentrations.

\section{Applications: Ice and other Materials}\label{sec:app}

Thermal regelation of particles can occur in terrestrial ice masses.   For example, ice sheets contain deposits of tephra particles from past volcanic eruptions\cite{NARCISI2005253}, dust particles originating from arid regions\cite{GROUSSET} and DNA remnants from ancient ecosystems\cite{Willers}. 
Accurate dating of ice core data, such as the isotopic composition, chemical species (in ice and on particles), provides the highest resolution reconstructions of past climates. 
Layers of volcanic particles are used as absolute time markers for dating the cores\cite{dunbar2003,NARCISI2006}. The particles are typically surrounded by impurities \cite{DUNBAR2011} and experience a temperature gradient in the lower--oldest and most compressed--part of the ice sheet due to the basal geothermal heat flux\cite{GUNDESTRUP1991,Fishere1500093}. Therefore, quantifying the translation of particles by thermal regelation and diffusion is important for the dating procedure. 

Because premelting occurs in many materials thermal regelation may be used, for example, as a novel method to tailor the properties of particle-reinforced composites\cite{zhang2005aligned,hanemann2010polymer}. During their manufacture, by adjusting the impurity concentration and the temperature gradient, the rate of thermal regelation of particles may be manipulated to achieve the desired distribution of particles. Indeed, premelting enhanced particle diffusion has been proposed as such a manipulation strategy\cite{peppin2009onsager}.

\section{Conclusions}\label{sec:conclusions}

Extending the ideas developed in the literature\cite{dash2006physics, dash1995premelting,JSWARFM,GILPIN1979235, worster1999fluid,wettlaufer1999,rempel2004premelting,hansen2010theory,peppin2009onsager, rempel2001interfacial} we have included the effects of impurities in the theory of 
thermomolecular pressure driven thermal regelation of spherical particles. In \S \ref{sec:theory} we provide the formulae for the speed and displacement of particles.  By combining these with their premelting facilitated diffusivity we derive a Fokker-Planck-like equation (Eq.\ref{equation17}) for the evolution of the probability density of particles, $f(\mathbf{x}, t)$, in the dilute limit.  
We illustrate the theory in \S \ref{sec:Results} by considering as a model system silicon particles in ice, which can either be considered as a transparent composite material for laboratory studies, or as a setting to examine the redistribution of particles in ice cores studies of paleoclimate.  We systematically examined the role of particle size, impurity concentration and temperature gradient.  Because of the known sensitivity of premelted film thickness on impurity concentration, for a given particle size we find that particle displacement is very sensitive to impurities and the temperature gradient, which underlies the thermomolecular pressure gradient.   As discussed in the analysis surrounding figure \ref{figure6}, these effects control the evolution of $f(\mathbf{x}, t)$  which can be diffusively or advectively dominated, of particular relevance for the control and manipulation of materials properties and quantifying the particle based time horizons in ice cores.  
Consequences of these findings are important for interpretation of ice core dating, which often relies on time calibration from volcanic particles, and provide a framework for the manipulation of particles in composite media.  In the former case, we found micron sized particles could be displaced tens of meters in thousands of years, compromising dating accuracy and in the latter case 
we found that silicon particles can be moved centimeters through ice on time scales of minutes.   

\section*{Conflicts of interest}
 There are no conflicts to declare.

\section*{Acknowledgements}
The authors acknowledge the support of the Swedish Research
Council Grant No. 638-2013-9243.




\section*{References}
 \providecommand*{\mcitethebibliography}{\thebibliography}
\csname @ifundefined\endcsname{endmcitethebibliography}
{\let\endmcitethebibliography\endthebibliography}{}

\end{document}